\documentclass[conference]{IEEEtran}
\IEEEoverridecommandlockouts

\usepackage{amsmath,amsfonts,amssymb,amsthm}
\theoremstyle{plain}

\theoremstyle{definition}

\theoremstyle{remark}

\usepackage{mathtools}
\usepackage{mathrsfs}

\usepackage{enumerate}
\usepackage{cite}
\usepackage{url}

\usepackage{geometry}
\newgeometry{top=1in,bottom=0.75in,right=0.75in,left=0.75in}

\usepackage{graphicx}
\graphicspath{ {Figures/} }
\usepackage{subfig}

\usepackage{tikz}
\usetikzlibrary{shapes, shapes.symbols, 
                decorations, decorations.markings, decorations.pathmorphing, 
                calc, arrows, backgrounds, positioning, fit, matrix}

\def \PF{\text{PF}}
\def \T{\top}
\def \ba{\arraycolsep=3pt\begin{array}}
\def \ea{\end{array}}
\def \be{\begin{equation}}
\def \ee{\end{equation}}

\def \bb{\mathbb}
\def \mc{\mathcal}
\def \ms{\mathsf}
\def \conv{\mathrm{conv}}
\def \nom{\text{nom}}
\def \rec{\text{rec}}
\def \Min{\text{minimize}}
\def \st{\text{subject to}}
\def \BL{\text{BL}}

\DeclareMathOperator*{\argmin}{arg\,min}

\begin{document}

\title{Incentive Design for Eco-driving in Urban Transportation Networks}

\author{M. Umar B. Niazi \and Jung-Hoon Cho \and Munther A. Dahleh \and Roy Dong \and Cathy Wu 
\thanks{M. U. B. Niazi and M. A. Dahleh are with the Laboratory for Information and Decision Systems, Department of Electrical Engineering and Computer Science, Massachusetts Institute of Technology, Cambridge, MA 02139, USA. Email: \texttt{niazi@mit.edu}, \texttt{dahleh@mit.edu}}
\thanks{J.-H. Cho and C. Wu are with the Laboratory for Information and Decision Systems, Department of Civil and Environmental Engineering, Massachusetts Institute of Technology, Cambridge, MA 02139, USA. Email: \texttt{jhooncho@mit.edu}, \texttt{cathywu@mit.edu}}
\thanks{R. Dong is with the Department of Industrial and Enterprise Systems Engineering, University of Illinois at Urbana-Champaign, Urbana, IL 61801, USA. Email: \texttt{roydong@illinois.edu}}
\thanks{This work was partially supported by the National Science Foundation (NSF) under grant number 2149548, the European Union's Horizon Research and Innovation Programme under Marie Sk\l{}odowska-Curie grant agreement No. 101062523, and the Kwanjeong scholarship.}
}

\maketitle

\begin{abstract}
    Eco-driving emerges as a cost-effective and efficient strategy to mitigate greenhouse gas emissions in urban transportation networks. Acknowledging the persuasive influence of incentives in shaping driver behavior, this paper presents the `eco-planner,' a digital platform devised to promote eco-driving practices in urban transportation. At the outset of their trips, users provide the platform with their trip details and travel time preferences, enabling the eco-planner to formulate personalized eco-driving recommendations and corresponding incentives, while adhering to its budgetary constraints. Upon trip completion, incentives are transferred to users who comply with the recommendations and effectively reduce their emissions. By comparing our proposed incentive mechanism with a baseline scheme that offers uniform incentives to all users, we demonstrate that our approach achieves superior emission reductions and increased user compliance with a smaller budget.
\end{abstract}

\section{Introduction}

In many countries, transportation accounts for a significant share of greenhouse gas emissions, ranging from a quarter to one-third of the total emissions. Numerous strategies are being employed to improve fuel efficiency and decrease emissions in on-road vehicles. These approaches encompass advancements in engine and vehicle technologies as well as improvements in fuel quality. However, eco-driving stands out as the most cost-effective and highly efficient means of reducing emissions from road transportation \cite{huang2018}. Research has consistently demonstrated that eco-driving practices can yield substantial reductions in vehicle emissions, ranging from 10\% to as high as 45\% \cite{sivak2012}. These findings underscore the immediate and significant role that eco-driving can play in addressing the challenge of climate change.

Eco-driving is a set of techniques that aim to improve fuel efficiency and reduce emissions by optimizing vehicle operation and driver behavior. Some of the techniques include:
\begin{itemize}
    \item Selecting less congested routes: This reduces fuel consumption and emissions by minimizing idling and stop-and-go traffic.
    \item Improving driving style: This includes avoiding aggressive acceleration and braking, and maintaining a steady speed. These techniques reduce fuel consumption and emissions by minimizing energy losses due to inefficient vehicle operation.
\end{itemize}
It is important to note that opting for less congested routes may result in longer travel times, and improving driving style might require adjustments that some drivers find inconvenient. To overcome these challenges and promote eco-driving, there may be a need for transportation system operators to introduce incentives that encourage individuals to adopt these practices for the purpose of emission reduction.

A considerable amount of evidence supports the effectiveness of incentive programs in promoting eco-driving. For instance, \cite{lai2015} observed a reduction of over 10\% in fuel consumption and emissions after monetary incentives were introduced to bus drivers. Remarkably, they found that the cost savings in fuel for bus companies exceeded the incentives provided to drivers. Comparable results were obtained by \cite{liimatainen2011} and \cite{schall2017} when incentivizing heavy-duty vehicle drivers in logistic companies. Furthermore, behavioral studies, such as \cite{mcconky2018} and \cite{vaezipour2019}, demonstrate that monetary incentives are more effective in changing driver behavior than providing informational content on eco-driving through in-vehicle human-machine interfaces.
Nevertheless, the incentive schemes presented in this body of literature are overly simplistic and do not cater to various driver types with differing preferences.

In this paper, we propose a digital platform that incentivizes human drivers to eco-drive with the goal of reducing the overall emissions of an urban transportation network. At the beginning of their trips, the users provide private information and preferences to the platform, such as their origin and destination, vehicle type, and preferred travel time vs. emissions trade-offs. Using this information, the platform computes feasible eco-routing and eco-driving strategies for each user. Then, it devises personalized incentives and eco-driving recommendations to users by minimizing overall emissions subject to budget constraints.

The overarching goal of the platform is to minimize the network's emissions while optimally allocating a limited budget as incentives to drivers. Our approach is unique in its integration of a traffic simulator into the incentive mechanism. The simulator predicts traffic conditions and calculates corresponding eco-driving recommendations and optimal incentives that can potentially reduce emissions by a certain amount. This feature allows us to account for real-time traffic variations in our incentive strategy.

The paper is organized as follows. Section~\ref{sec:incentive-mechanism} introduces the incentive mechanism for eco-driving. Section~\ref{sec:technical} delves into some technical observations and remarks. Section~\ref{sec:experiments} presents numerical experiments. Finally, Section~\ref{sec:conclusion} summarizes the findings and highlights potential future directions.

\section{Incentive Mechanism for Eco-driving}
\label{sec:incentive-mechanism}

In this section, we present a model of incentive mechanism for drivers in a transportation network. Drivers are assumed to be cost minimizers who choose an optimal outcome over their feasible sets in terms of emissions and travel time. We propose a method for computing these feasible sets using a microsimulator for traffic.

\subsection{Model Setup and Assumptions}

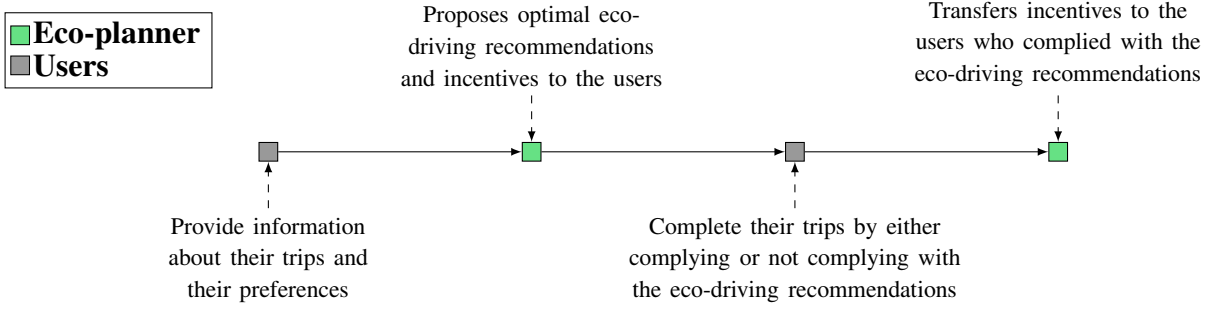
\begin{figure*}[!t]
    \centering
    \begin{tikzpicture}
        \node[regular polygon, regular polygon sides=4, draw, inner sep=0pt, minimum size=10pt, fill=black!40] at (-3.3,1.9) {};
        \node[anchor=west] at (-3.25,1.9) {\large \textbf{Users}};

        \node[regular polygon, regular polygon sides=4, draw, inner sep=0pt, minimum size=10pt, fill=blue!20!green!60] at (-3.3,2.3) {};
        \node[anchor=west] at (-3.25,2.3) {\large \textbf{Eco-planner}};

        \draw (-3.5,2.6) rectangle (-0.75,1.6);

        \node[regular polygon, regular polygon sides=4, draw, inner sep=0pt, minimum size=10pt, fill=black!40] (step1) at (0,0.75) {};
        \draw[-latex, dashed] (0,0) -- (step1);
        \node[anchor=north, text width=3cm, align=center] at (0,0) {\small Provide information about their trips and their preferences};
        
        \node[regular polygon, regular polygon sides=4, draw, inner sep=0pt, minimum size=10pt, fill=blue!20!green!60] (step2) at (3.5,0.75) {};
        \draw[-latex, dashed] (3.5,1.5) -- (step2);
        \node[anchor=south, text width=5cm, align=center] at (3.5,1.5) {\small Proposes optimal eco-driving recommendations and incentives to the users};
        
        \node[regular polygon, regular polygon sides=4, draw, inner sep=0pt, minimum size=10pt, fill=black!40] (step3) at (7,0.75) {};
        \draw[-latex, dashed] (7,0) -- (step3);
        \node[anchor=north, text width=5cm, align=center] at (7,0) {\small Complete their trips by either complying or not complying with the eco-driving recommendations};
        
        \node[regular polygon, regular polygon sides=4, draw, inner sep=0pt, minimum size=10pt, fill=blue!20!green!60] (step4) at (10.5,0.75) {};
        \draw[-latex, dashed] (10.5,1.5) -- (step4);
        \node[anchor=south, text width=4cm, align=center] at (10.5,1.5) {\small Transfers incentives to the users who complied with the eco-driving recommendations};
        
        \path[-latex]
        (step1) edge (step2)
        (step2) edge (step3)
        (step3) edge (step4);
    \end{tikzpicture}
    \caption{Steps in the proposed incentive mechanism (from left to right).}
    \label{fig:incentive-mechanism}
\end{figure*}

We consider an urban transportation network denoted by a graph $\mc{N}=(\mc{I},\mc{L})$, where $\mc{I}$ denotes the set of intersections and $\mc{L}\subseteq\mc{I}\times\mc{I}$ denotes the set of links/roads. We aim to design an eco-planner digital platform $\ms{P}$ that plans and recommends eco-driving strategies to its users in the network $\mc{N}$ before they embark on their trips. The eco-planner $\ms{P}$ also commits to providing certain incentives if the users comply with the eco-driving recommendations to reduce their emissions at the expense of increased travel time. The users who subscribe to $\ms{P}$ are denoted by a set $U=\{1,\dots,n\}$, where $n$ denotes the number of users.

Before starting her commute, each user~$i$, $i\in U$, asks $\ms{P}$ to plan the journey by providing her private information, which includes the following:
\begin{itemize}
    \item Origin-Destination pair $(o_i,d_i)\in\mc{L}\times\mc{L}$, which corresponds to the start and end points of $i$'s journey
    \item Vehicle type $v_i\in V$, where $V=\{v_1,\dots,v_l\}$ is the set of finite number of vehicle types
    \item Emissions vs. travel time trade-off parameter $\vartheta_i\in[0,1]$.
\end{itemize}
For instance, vehicle type may correspond to not only its classification (sedan, SUV, etc.) but also its engine and fuel types. 
This information is needed so that $\ms{P}$ can predict the emissions of user~$i$ on her commute between $(o_i,d_i)$ on different routes with different traffic conditions. Emissions vs. travel time trade-off parameter $\vartheta_i$ is needed to assess the urgency of user~$i$ for her trip, and whether she will accept a certain amount of incentive to reduce her emissions by eco-driving by compromising slightly on her travel time. In this paper, we assume that each user~$i$ provides her information $(o_i,d_i,v_i,\vartheta_i)$ truthfully.

After obtaining the private information $(o_i,d_i,v_i,\vartheta_i)_{i\in U}$, the eco-planner $\ms{P}$ predicts the best eco-driving strategies in terms of a route choice and driving style for each user depending on the predicted traffic conditions for their trips. Then, $\ms{P}$ recommends those strategies and persuades the users to follow those strategies by offering incentives $\gamma_1,\dots,\gamma_n$ subject to $\ms{P}$'s budget constraint $\sum_{i=1}^n\gamma_i\leq B$, where $B\in\bb{R}_{>0}$ is the total budget of $\ms{P}$. The users complete their trips by either complying or not complying with the eco-driving recommendations, and their commutes result in a certain outcome in terms of their individual emissions. Finally, $\ms{P}$ transfers the incentives to the users based on their compliance with the recommendations and to compensate for their increased travel times due to their eco-driving and reducing emissions. The incentive mechanism is summarized in Fig.~\ref{fig:incentive-mechanism}.

As stated earlier, we assume that users are truthful and do not strategically manipulate their information. This avoids the issue of adverse     selection. Incorporating incentive compatibility constraints will be a topic of our future work. Secondly, to address the issue of moral hazard, whereby the behavior and outcomes of users are not observable, we assume $\ms{P}$ employs vehicle telematics \cite{handel2014, young2020} to measure the driving style and emissions of each user during her commute. Finally, we assume traffic conditions and network structures where eco-driving results in lower emissions and longer travel times. For example, choosing a longer route with fewer intersections would yield lower emissions as compared to choosing a shorter route through an urban area with plenty of intersections and stop signs. The former route may result in longer travel times, but because of smooth driving, it would result in lower emissions than the latter route. It is important to remark that there could be other scenarios, e.g., eco-driving near signalized intersections \cite{jayawardana2022}, where eco-driving improves both travel times and emissions. However, in these scenarios, user's and eco-planner's objectives are aligned, and the issue becomes one of information design rather than incentive design.

\subsection{Users' Objectives}

Since eco-driving strategies can increase travel time by requiring drivers to take longer routes or drive at slower speeds, we assume that drivers, being cost minimizers, optimize their driving behaviors based on the utilities they place on reducing their emissions/fuel consumption versus reducing their travel times.
Let $x_i=[\ba{cc} x_i^t & x_i^e \ea]^\T \in X_i \subset \bb{R}_{>0}^2$ denote the outcome of user~$i$'s commute between the origin $o_i$ and the destination $d_i$, where $x_i^e$ denotes her emissions and $x_i^t$ denotes her travel time. Here, $X_i$ is assumed to be a convex set denoting all feasible emission-travel time pairs achievable via different driving styles and route choices for $(o_i,d_i)$ given $i$'s vehicle type $v_i$ and traffic conditions in the network $\mc{N}$. In other words, each point $x_i\in X_i$ denotes an emissions-travel time outcome corresponding to a certain route and driving style. 

Each user~$i$ chooses her route and driving style that minimizes her cost function 
\[
c_i(x_i)=(1-\vartheta_i)x_i^t + \vartheta_i x_i^e
\]
where $\vartheta_i\in[0,1]$ is user~$i$'s emissions vs. travel time trade-off parameter.
In other words, $i$ solves the following convex optimization problem:
\begin{equation}
    \Min~ c_i(x_i) \triangleq \theta_i^\T x_i ~\st~ x_i\in X_i
    \label{eq:min_i}
\end{equation}
where $\theta_i = [\ba{cc} 1-\vartheta_i & \vartheta_i \ea]^\T$ is $i$'s preference parameter.
Let
\begin{equation}
    x_i^{\nom} = \argmin_{x_i\in X_i} c_i(x_i)
    \label{eq:argmin_i}
\end{equation}
be the nominal outcome of user~$i$.

\subsection{Computation of Feasible Sets}

Let $R_i$ denote the set of routes for the origin-destination pair $(o_i,d_i)$, where $r_{i,j}\in R_i$ is the $j$-th route containing a path in $\mc{N}$ starting from $o_i$ and ending at $d_i$.
The eco-planner invokes a traffic simulator $\ms{S}$ to compute the feasible sets of each user. In other words, the simulator $\ms{S}$ takes the network $\mc{N}$, origin-destination pair $(o_i,d_i)$, and vehicle type $v_i$ as inputs, and outputs the feasible set $X_i\subset\bb{R}_{>0}^2$. 

The first step of the simulator involves computation of $p$ shortest routes $R_i=(r_{i,1},\dots,r_{i,p})$ in $\mc{N}$ between $(o_i,d_i)$. After that, for every $j\in[p]$, $\ms{S}$ computes a set of $m\in\bb{N}$ points $\hat{X}_{ij}=\{x_{i,j}^1,\dots,x_{i,j}^m\}$, which results from simulating different eco-driving as well as normal driving styles on route $r_{i,j}$ under different traffic conditions. Finally, $\ms{S}$ outputs
\begin{equation}
    \label{eq:conv_feas_X_i}
    X_i=\conv(\hat{X}_{i1},\dots,\hat{X}_{ip})
\end{equation}
where $\conv(\cdot)$ denotes the convex hull. Notice that computing feasible sets as in \eqref{eq:conv_feas_X_i} renders \eqref{eq:min_i} to be a linear program.

\subsection{Incentive Mechanisms} \label{subsec_incentive}

Before starting their trips, users interact with the eco-planner by providing their information (see Fig.~\ref{fig:incentive-mechanism}). The eco-planner then proposes incentives $\gamma_1,\dots,\gamma_n$ to the users conditioned on their following corresponding eco-driving strategies to reduce emissions. 
Here, $\gamma_i\in\bb{R}_{\geq 0}$ denotes the incentive user~$i$ receives from the eco-planner~$\ms{P}$ at the end of her trip between $(o_i,d_i)$ if she followed the recommended route and eco-driving strategies to achieve the recommended outcome $x_i^\rec\in X_i$.

\subsubsection{Baseline Incentive Mechanism}
In this paper, we consider the baseline incentive mechanism as equally allocating the total budget among all users of the same type, where we do not consider their parameters $\vartheta_i$'s. For instance, if all the users have the same origin-destination pairs $(o_i,d_i)$ and the same vehicle types $v_i$, then each user~$i$ is offered the same incentive $\gamma_i^{\BL}=B/n$ conditioned on achieving the recommended outcome $x_i^\rec$. In this paper, we use this baseline mechanism to compare with our proposed optimal mechanism. In our future work, other baselines will also be devised for comparison.

\subsubsection{Optimal Incentive Mechanism} 
To compute the optimal eco-driving recommendations and corresponding incentives for each user while adhering to budget constraints, the proposed incentive mechanism (Fig.~\ref{fig:incentive-mechanism}) involves the following steps:
\begin{enumerate}
    \item Users report their information $(o_i,d_i,v_i,\vartheta_i)_{i=1,\dots,n}$ to the eco-planner $\ms{P}$.
    \item $\ms{P}$ invokes the simulator $\ms{S}$ and obtains 
    \[
    X_i=\ms{S}(\mc{N},o_i,d_i,v_i), \; \forall i\in[n].
    \]
    \item Using $X_i$, $\ms{P}$ solves \eqref{eq:min_i} and finds $x_i^\nom$ from \eqref{eq:argmin_i}.
    \item $\ms{P}$ solves the following linear program:
    \begin{subequations}
    \begin{align}
        \Min~~  & \xi^\T x^\rec \\
        \st~~ & c_i(x_i^\rec) - \gamma_i \leq c_i(x_i^\nom) \\
        & \sum_{i=1}^n \gamma_i \leq B, \gamma_i\geq 0 \label{eq:budget-constraint} \\         
        & x_i^\rec\in X_i, ~\forall i\in[n]
    \end{align}
    \label{prob:eco-planner}%
    \end{subequations}
    where $x^\rec=[\ba{ccc} x_1^\rec & \dots & x_n^\rec \ea]\in X_1\times\dots\times X_n\subset\bb{R}_{>0}^{2n}$ is a vector of outcomes recommended by $\ms{P}$, $B$ is the total budget of $\ms{P}$, and
    \[
    \xi^\T = 1_n^\T \otimes [\ba{cc} 0 & 1 \ea]=[\ba{ccccc} 0 & 1 & \dots & 0 & 1 \ea]
    \]
    is a vector indicating that the goal of $\ms{P}$ is to minimize the total emissions of all users subject to the budget constraint \eqref{eq:budget-constraint}.
    \item $\ms{P}$ proposes incentives $\gamma_1,\dots,\gamma_n$ to the users conditioned on complying with the recommended routes and eco-driving strategies that yield the emission-travel time outcomes $x_1^\rec,\dots,x_n^\rec$, respectively.
\end{enumerate}

Notice that the linear program \eqref{prob:eco-planner} can be written in the standard form by choosing
\begin{equation}
    \label{eq:incentive_gamma_i}
    \gamma_i = \theta_i^\T (x_i^\rec - x_i^\nom)
\end{equation}
and rewriting the convex polytope $X_i$ in \eqref{eq:conv_feas_X_i} as
\begin{equation}
    X_i=\{x_i\in\mathbb{R}_{> 0}^2: A_ix_i\leq b_i\}
\end{equation}
where $A_i\in\bb{R}^{k\times 2}$ and $b_i\in\bb{R}^{k}$ with $k\leq mp$. Here, $p$ corresponds to the number of routes chosen between the origin $o_i$ and the destination $d_i$, and $m$ is the number of different driving styles simulated on every route by $\ms{S}$. Thus, we can equivalently write \eqref{prob:eco-planner} as
\begin{subequations}
\begin{align}
    \Min~~  & \xi^\T x^\rec \\
    \st~~ & \mc{A}x^\rec\leq \beta
\end{align}
\label{prob:eco-planner2}
\end{subequations}
where
\[
\mc{A} = \left[\ba{ccc} 
\theta_1 & \dots & \theta_n \\
A_1 & & \\
& \ddots & \\
& & A_n
\ea\right], \quad \beta = \left[\ba{c} 
B + \theta^\T x^\nom \\
b_1 \\
\vdots \\
b_n
\ea\right]
\]
with $\theta^\T=[\ba{ccc} \theta_1 & \dots & \theta_n \ea]$ and $x^\nom = [\ba{ccc} x_1^\nom & \dots & x_n^\nom \ea]^\T$. 

\section{Technical Observations and Remarks}
\label{sec:technical}

\subsection{Computed Feasible Sets are only Approximations}

We remark that the feasible sets obtained from the simulator are only approximations of the true feasible sets of the users, which can be refined by simulating more number $m$ of traffic scenarios and driving styles. However, using approximated feasible sets does not significantly impact the incentive mechanism. Depending on the quality of the approximation, the eco-planner $\ms{P}$ can relax the terms of the contract by transferring the incentive to user~$i$ if her actual outcome $x_i$ at the end of the trip turned out to be inside a ball of certain radius around the recommended outcome $x_i^\rec$.

To elucidate, the eco-planner $\ms{P}$ computes an approximated feasible set $\hat{X}_i\subseteq X_i$ from the simulator $\ms{S}$, $\hat{X}_i$ is a convex polytope. The inclusion of $\hat{X}_i$ inside $X_i$ is because $X_i$ is convex and no simulated point can be outside of $X_i$. In the case where $\ms{S}$ is able to simulate the boundary points of $X_i$, $\hat{X}_i$ will be a convex polytope inscribed inside $X_i$. Nevertheless, the eco-planner $\ms{P}$ will compute the nominal solution $\hat{x}_i^\nom\in\hat{X}_i$ to \eqref{eq:argmin_i}, which will be a projection of the true $x_i^\nom\in X_i$ onto $\hat{X}_i$. Putting computational issues aside, if the simulator $\ms{S}$ simulates a very large number of cases $m$, the approximation $\hat{X}_i$ can be arbitrarily improved. Keeping this in mind, $\ms{P}$ transfers the incentive if the actual outcome $x_i\in X_i$ is with an $\varepsilon$-Euclidean distance from the recommended outcome $x_i^\rec\in\hat{X}_i$.

\subsection{Nominal Outcomes are on the Pareto Fronts}

It is important to note that the nominal solution $x_i^\nom$ of user~$i$ lies at a certain part of the boundary of her feasible set $X_i$. To explain this fact, we define the following notions.
We say $x_i=[\ba{cc} x_i^t & x_i^e \ea]^\T \in X_i$ \textit{pareto dominates} $\Tilde{x}_i=[\ba{cc} \Tilde{x}_i^t & \Tilde{x}_i^e \ea]^\T \in X_i$ if either (i)~$x_i^t \leq \Tilde{x}_i^t$ and $x_i^e<\Tilde{x}_i^e$ OR (ii)~$x_i^t < \Tilde{x}_i^t$ and $x_i^e \leq \Tilde{x}_i^e$.
Moreover, $x_i$ is said to be \textit{pareto optimal} if there does not exist $\Tilde{x}_i\in X_i$ that pareto dominates $x_i$. Then, the \textit{pareto front} of the feasible $X_i$ is defined as
\[
    \PF(X_i)=\{x_i\in X_i : x_i ~\text{is pareto optimal}\}
\]
which will be a subset of $X_i$'s boundary. 

In light of the above, we have $x_i^\nom\in\PF(X_i)$, where $x_i^\nom$ is the solution of the convex problem \eqref{eq:argmin_i}. To prove this claim, assume $x_i^\nom\notin\PF(X_i)$. Then, $x_i^\nom$ is not pareto optimal and there exists $v\in\bb{R}_{\geq 0}^2$, $v\neq 0$, such that $x_i^\nom-v$ pareto dominates $x_i^\nom$ and
\[
c_i(x_i^\nom) = \theta_i^\T x_i^\nom > \theta_i^\T (x_i^\nom - v) = c_i(x_i^\nom-v)
\]
which is a contradiction becausue $x_i^\nom$ being the solution of \eqref{eq:argmin_i} is the minimizer of $c_i(\cdot)$.

\subsection{Users can Report their Preferred Travel Times}

Some might argue that reporting the emissions vs. travel time trade-off parameter, $\vartheta_i$, can be challenging for users in reality. Aside from its vague and technical interpretation, it is possible that users may not know the exact value of their parameters, let alone report them truthfully. Instead, the eco-planner may ask the user to report their preferred (i.e., nominal) travel time $x_i^{t,\nom}\in X_i$ between $o_i$ and $d_i$. Notice that this report has to correspond to achievable travel time on $(o_i,d_i)$ at the time of report. Since we know $x_i^\nom\in\PF(X_i)$, we can find the corresponding emissions $x_i^{e,\nom}$ that $i$ would incur for their nominal route selection and driving style. Then, the question is how to estimate $\theta_i=[\ba{cc} \vartheta_i & 1-\vartheta_i\ea]^\T$ given that $x_i^\nom$ is the minimizer of \eqref{eq:min_i}?

A simple algorithm to estimate $\theta_i$ is as follows. Sample $\vartheta_i\in[0,1]$ and obtain $0=\vartheta_i^1 < \vartheta_i^2 < \dots < \vartheta_i^s=1$ for some sufficiently large $s\in\bb{N}$. For every $\vartheta_i^j$, solve \eqref{eq:argmin_i} and obtain the solution $x_i^{\PF_j}\in\PF(X_i)$. Then, 
\[
s^* = \argmin_{j\in\{1,\dots,s\}} \|x_i^\nom - x_i^{\PF_j}\|
\]
and
\be 
\label{eq:hat_theta}
\hat{\theta}_i = \left[\ba{c} \vartheta_i^{s^*} \\ 1-\vartheta_i^{s^*} \ea\right].
\ee
One can refine this solution arbitrarily by subsequently sampling again around $x_i^{\PF_{s^*}}$, where the idea is to coarsely sample initially and then keep on refining the samples around the closest solutions in the next iteration.

When users report their preferred travel times instead of their emissions vs. travel time trade-off parameters, the steps involved in the incentive mechanism can be modified accordingly and the linear program \eqref{prob:eco-planner2} is written as
 \begin{subequations}
    \begin{align}
        \Min~~  & \xi^\T x^\rec \\
        \st~~ & \hat{\mc{A}}x^\rec\leq \hat{\beta}
    \end{align}
    \label{prob:eco-planner3}
\end{subequations}
where 
\[
    \hat{\mc{A}} = \left[\ba{ccc} 
    \hat{\theta}_1 & \dots & \hat{\theta}_n \\
    \hat{A}_1 & & \\
    & \ddots & \\
    & & \hat{A}_n
    \ea\right], \quad \hat{\beta} = \left[\ba{c} 
    B + \hat{\theta}^\T x^\nom \\
    \hat{b}_1 \\
    \vdots \\
    \hat{b}_n
    \ea\right]
\]
with $\hat{\theta}^\T=[\ba{ccc} \hat{\theta}_1 & \dots & \hat{\theta}_n \ea]$. $\ms{P}$ then proposes incentives $\gamma_1,\dots,\gamma_n$ to the users, where 
\[
    \gamma_i=\hat{\theta}_i^\T (x_i^\rec-x_i^\nom).
\]
The incentives are conditioned on the recommended emission-travel time outcomes $x_1^\rec,\dots,x_n^\rec$, respectively, which are achieved by complying with the recommended routes and eco-driving strategies

\section{Experiments}
\label{sec:experiments}

\subsection{Road Network}

\begin{figure}[!t]
    \centering
    \begin{tikzpicture}
        \node[regular polygon, regular polygon sides=3, draw, inner sep=0pt, minimum size=6pt, fill=black] (o1) at (-1,0) {};
        \node[circle, draw, inner sep=0pt, minimum size=6pt] (o) at (0,0) {};
        \draw[-latex] (o1) -- (o);
        \node at (-0.5,-0.4) {\small Origin};
        \node[circle, draw, inner sep=0pt, minimum size=6pt] (d) at (6,0) {};
        \node[regular polygon, regular polygon sides=3, draw, inner sep=0pt, minimum size=6pt, fill=black] (d1) at (7,0) {};
        \draw[-latex] (d) -- (d1);
        \node at (6.5,-0.4) {\small Destination};
        
        \node[circle, draw, inner sep=0pt, minimum size=6pt] (s1) at (1.5,0) {};
        \node[circle, draw, inner sep=0pt, minimum size=6pt] (s2) at (4.5,0) {};
        \node at (1.5,-0.4) {\includegraphics[width=0.025\textwidth]{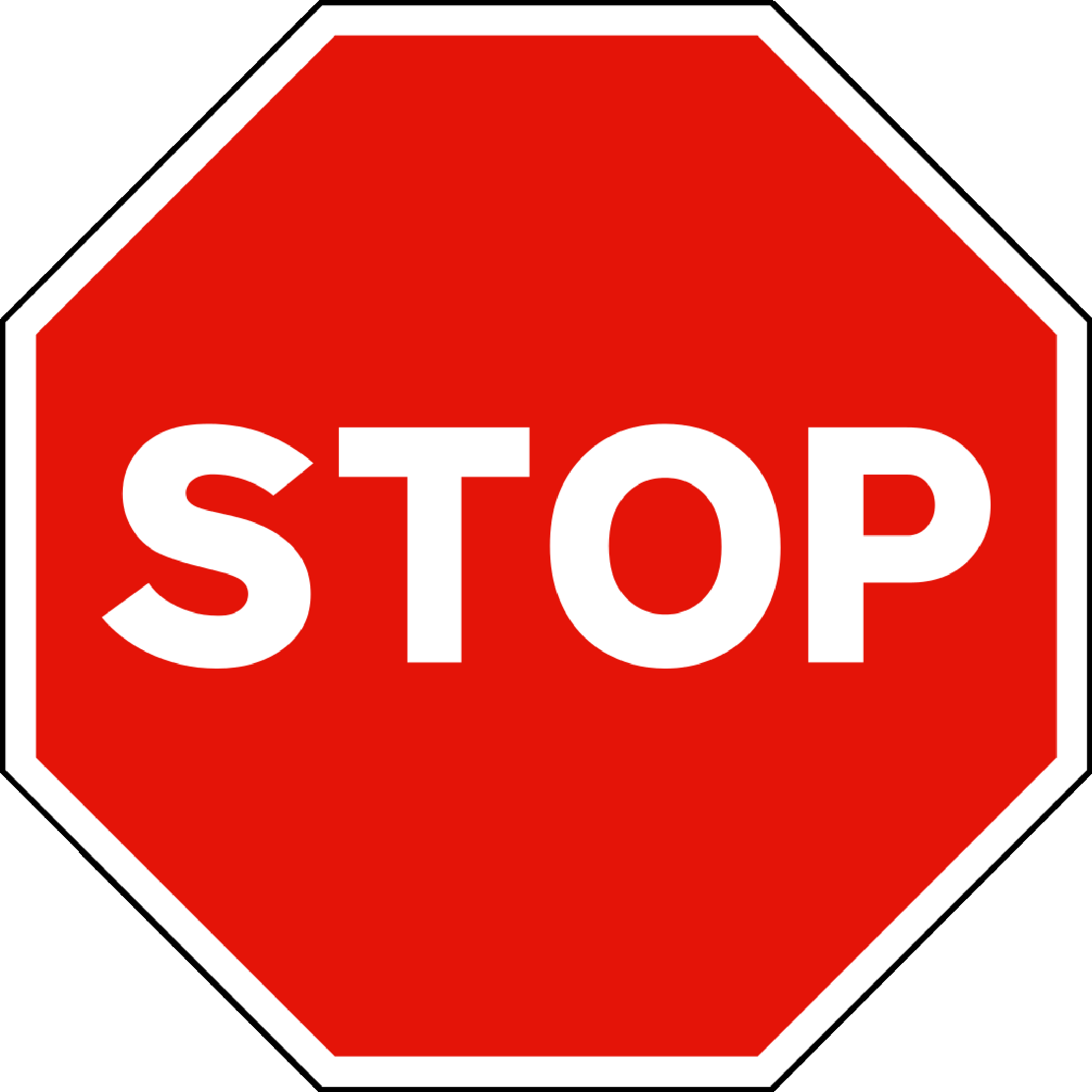}};
        \node at (4.5,-0.4) {\includegraphics[width=0.025\textwidth]{Figures/traffic-stop.eps}};
        \node[circle, draw, inner sep=0pt, minimum size=6pt] (s3) at (3,0) {};
        \node at (3,-0.43) {\includegraphics[width=0.03\textwidth]{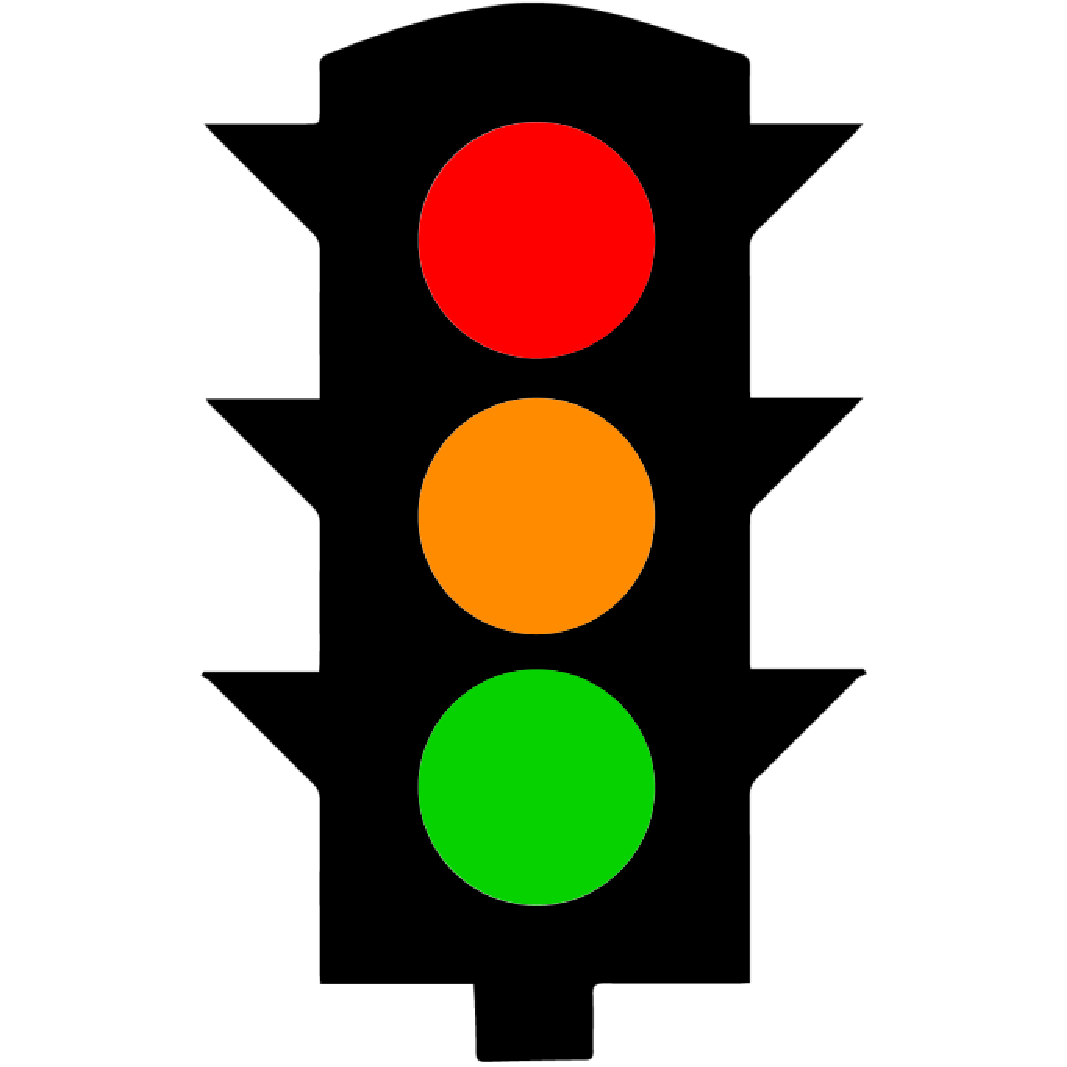}};
        \path[-latex, thick]
        (o) edge (s1)
        (s1) edge (s3)
        (s3) edge (s2)
        (s2) edge (d)
        (o) edge[bend left=50] node[pos=0.5, above] {\scriptsize Route 2 (Longer, via a boulevard)}(d);
        \node at (3,0.4) {\scriptsize Route 1 (Shorter, via an urban area)};
    \end{tikzpicture}
    \caption{Illustration of the road network with two alternative routes used in the simulation experiments.}
    \label{fig:road-network}
\end{figure}

We set up a controlled experiment with a simple road network using the Simulation of Urban MObility (SUMO) \cite{SUMO2018}. The road network consists of two routes for one origin-destination pair: the first route through the urban area is shorter but includes two stop signs, while the second route using a boulevard/highway, although longer, is uninterrupted by stop signs (Fig.~\ref{fig:road-network}). The speed limit for both routes is set to be the same. The first route, with two stop signs and one static traffic signal, is shorter and takes shorter travel time, but it emits larger amounts of carbon emissions because of stop-and-go traffic behavior. The second route is longer and may take longer travel time, but it is more eco-friendly because of a smoother traffic. 

\subsection{Experiment Design}
Our experiment design is structured to analyze the trade-offs between travel time and carbon emissions, directly informed by the route choice for a given OD pair. 
We computationally generated free-flow traffic conditions for both routes, ensuring an unbiased assessment of their inherent characteristics. 
All vehicles follow the Intelligent Driver Model (IDM), which is a time-continuous car-following model proposed by \cite{treiber2000}. 
The emission model used in the simulation is based on the Handbook Emission Factors for Road Transport (HBEFA).
Route 1, while shorter, was observed to produce higher CO emissions because of the stops imposed by the stop signs and a traffic signal. 
In contrast, Route 2, despite its longer distance, resulted in lower emissions by benefiting from a continuous driving flow without interruptions.

\begin{figure}[!t]
    \centering
    \includegraphics[width=0.4\textwidth]{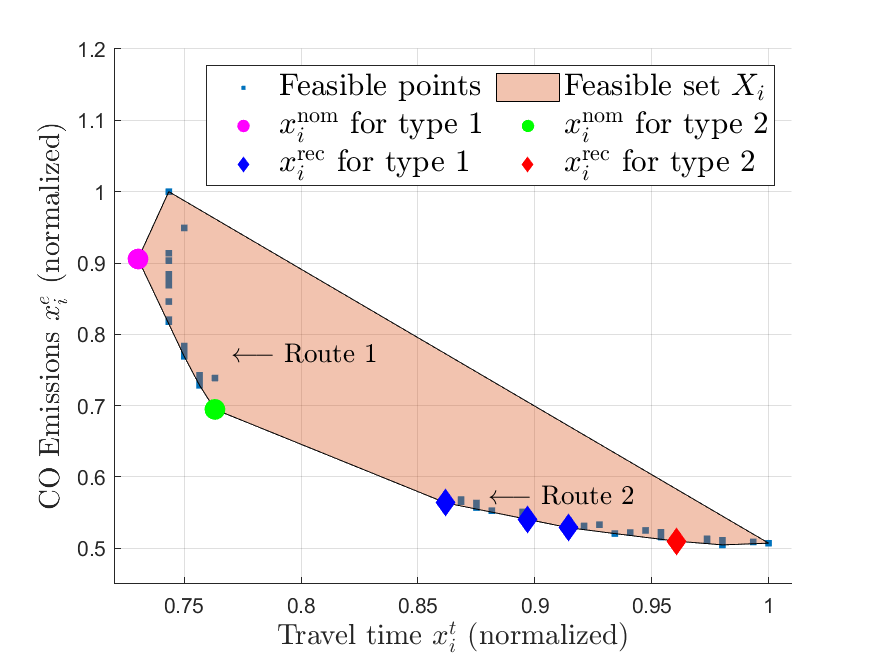}
    \caption{The feasible set computed from the feasible outcome points of both route~1 and 2. The nominal and recommended outcome points for the two driver types are indicated by circles and diamonds.}
    \label{fig:convex-hull}
\end{figure}

The collected data points, representative of the distinct travel times and emissions outcomes for each route, helped construct a convex hull that represents the feasible region of outcomes (Fig.~\ref{fig:convex-hull}). This convex hull is instrumental in visualizing the potential impact of different driving behaviors and route selections on travel time and carbon monoxide emissions. 
For each type, the nominal (circle) and recommended (diamond) outcome points are distinctly marked, providing insight into the specific eco-driving recommendations by the incentive mechanism.
This visual representation underscores the emission-saving potential of Route 2 despite its longer travel times, aligning with the eco-driving principles encouraged by the study's incentive mechanism.
Our approach allows us to evaluate the effectiveness of the proposed incentive mechanism in guiding drivers towards choices that align better with eco-friendly driving principles.

As described in Section~\ref{subsec_incentive}, we consider two incentive mechanisms: baseline and optimal (proposed). In the baseline, we propose equal incentive $\gamma_i^{\BL} = B/n$ to all the users. Let $x_i^{\rec,r2}\in X_i$ be a recommendation to users that corresponds to the least travel time on feasible points corresponding to route 2. Then, user~$i$ complies with this recommendation under baseline incentive if and only if
\[
\gamma_i^\BL \geq \theta_i^\T (x_i^{\rec,r2} - x_i^\nom).
\]
Similarly, given budget $B$, the optimal incentive mechanism yields $\gamma_1,\dots,\gamma_n$, where the user~$i$ complies with the recommendation $x_i^{\rec,r2}$ if and only if
\[
\gamma_i \geq \theta_i^\T (x_i^{\rec,r2} - x_i^\nom).
\]

\subsection{Experimental Results}
\begin{figure}[!t]
    \centering
    \subfloat[Compliance \label{fig:compliance-budget}]{\includegraphics[width=0.22\textwidth]{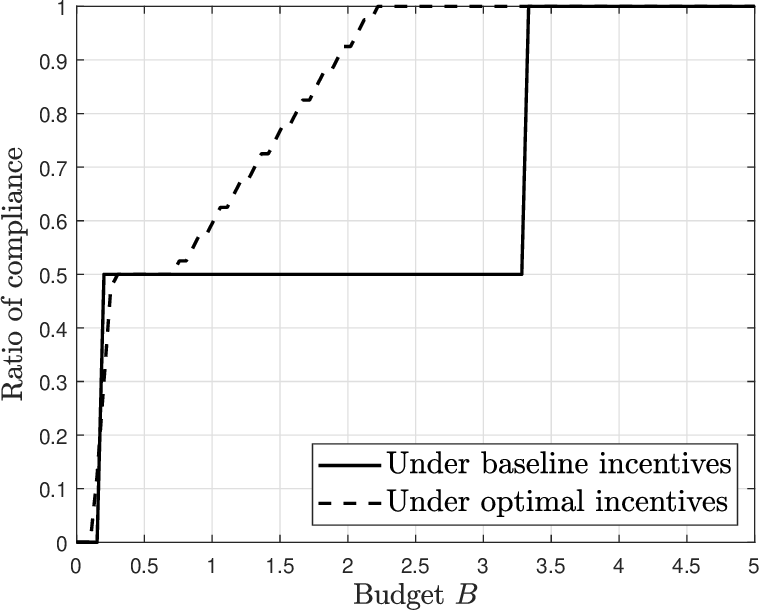}}
    \hfill
    \subfloat[Emissions and travel time \label{fig:emission-time-budget}]{
        \includegraphics[width=0.25\textwidth]{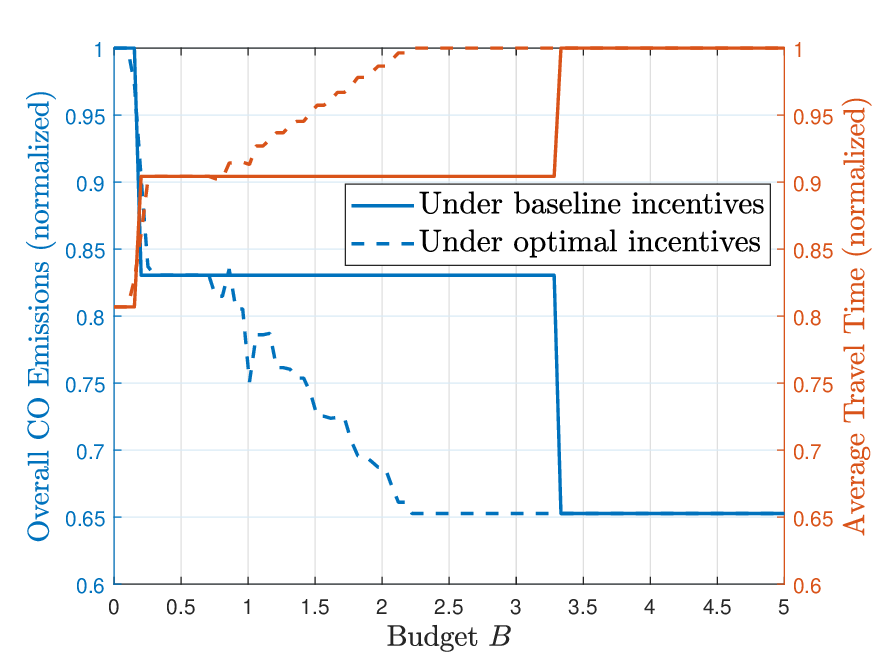}}
    \caption{Incentive mechanism under different levels of budget.}
    \label{fig:mech-budget}
\end{figure}

In Fig.~\ref{fig:compliance-budget}, we illustrate the correlation between budget allocation and driver compliance with eco-driving recommendations. It compares the result of the linear program in (\ref{prob:eco-planner3}) and baseline incentive across different budget levels. 
It is apparent that under baseline incentives, compliance escalates swiftly with a slight increase in budget, plateauing at a compliance ratio of 0.5. 
Beyond this point, the compliance rate remains constant, indicating that additional incentives under this model do not further motivate drivers. 
In contrast, under optimal incentives, we observe a gradual yet consistent increase in compliance as the budget grows, eventually surpassing the baseline once half of the drivers are compliant. 
This trend suggests that the optimal incentive structure is more effective at progressively encouraging a larger proportion of drivers to adopt eco-driving practices, particularly in scenarios where the budget is ample enough to support such incentivization.

Fig.~\ref{fig:emission-time-budget} illustrates the relationship between the budget allocated for incentives and the total CO emissions and average travel time from vehicles within the simulation. Both are normalized for comparative purposes. 
The graph compares the efficacy of baseline incentives to that of the proposed optimal incentives.
As shown in Fig.~\ref{fig:emission-time-budget}, both incentive schemes initially cause a sharp decrease in emissions as the budget is increased, demonstrating that even minimal financial motivation can significantly alter driving behaviors. However, the baseline incentives exhibit a more extended plateau compared to the optimal incentives, indicating a point where additional funds cease to yield proportional reductions in emissions. Conversely, the optimal incentives lead to a more consistent and prolonged decline in emissions with increasing budgets, highlighting their effectiveness in continuously promoting eco-friendly driving practices. Meanwhile, Fig.~\ref{fig:emission-time-budget} reveals an increase in average travel time concurrent with efforts to reduce carbon emissions, emphasizing the importance of a balanced strategy that judiciously weighs budget spending against both environmental impact and time-savings.

\section{Conclusion and Future Directions}
\label{sec:conclusion}

To promote eco-driving in urban transportation networks, we have developed an incentive mechanism in the form of a digital platform called eco-planner. Before starting their trips, users report their origins, destinations, vehicle types, and preferences for emissions versus travel time trade-offs. The eco-planner then simulates different traffic conditions and driving styles to generate optimal eco-driving recommendations and incentives for each user, subject to a budget constraint. In other words, eco-planner guarantees certain incentives to users who follow its recommendations. Users decide whether to comply with the recommendations based on the incentives offered. At the end of each trip, the eco-planner transfers the incentives to users who comply with the recommendations and reduce their emissions.

Several features of the incentive mechanism are noteworthy. First, the simulator efficiently computes the feasible outcomes for each user, helping the eco-planner to optimally choose incentives for all users. Second, the nominal outcomes of the users lie on the Pareto front of their feasible sets, enabling the platform to compute each user's incentive exactly using a simple expression \eqref{eq:incentive_gamma_i} depending on the corresponding recommendation.

Our future work includes modifying the incentive mechanism to be incentive-compatible, meaning that strategic users cannot maximize their rewards by reporting false information. Additionally, we will consider the effects of non-participating traffic on eco-driving of the participating users. In this scenario, the feasible sets computed by the platform may differ significantly from the actual feasible sets of the users, due to different traffic conditions and/or coupling between the driving behavior of users who share the same routes. By addressing these challenges, we can improve our incentive mechanism and make it more effective and suitable for deploying on real transportation networks.

\bibliographystyle{IEEEtran}
\bibliography{references}

\end{document}